\documentclass{nature}

\title{A More Complex Than Expected Formation History of the Milky Way's Last Major Merger}

\usepackage{graphicx}	% Including figure files
\usepackage{amsmath}	% Advanced maths commands
\usepackage{amssymb}	% Extra maths symbols 
\usepackage{float}

\usepackage{hyperref}
\hypersetup{colorlinks, citecolor=blue, filecolor=black, linkcolor=blue, urlcolor=blue }

\author{Hai-Feng~Wang$^{1,2, \dagger}$\thanks{Corresponding author: \texttt{haifeng.wang.astro@gmail.com; hfwang@bao.ac.cn}}, Guan-Yu~Wang$^{2, \dagger}$, Giovanni Carraro$^{2}$,  Gra\v{z}ina Tautvai\v{s}ien\.{e}$^{3}$, Joss Bland-Hawthorn$^{4,5}$, Thor Tepper-Garc\'ia$^{4,5}$}

\begin{document}

%\linenumbers

\maketitle

\begin{affiliations}
 \item  Local Universe and Time-Domain Astronomy Laboratory, Department of Astronomy, China West Normal University, Nanchong 637002, China 
 \item Dipartimento di Fisica e Astronomia ``Galileo Galilei", Universit\'a degli Studi di Padova, Vicolo Osservatorio 3, I-35122, Padova, Italy
 \item Institute of Theoretical Physics and Astronomy, Vilnius University, Saul\.{e}tekio av. 3, LT-10257 Vilnius, Lithuania
 \item Sydney Institute for Astronomy, School of Physics, University of Sydney, NSW 2006, Australia
 \item Centre of Excellence for All-Sky Astrophysics in Three Dimensions (ASTRO-3D), Australia
 \item[$^\dagger$] These authors contributed equally to this work.
\end{affiliations}

\begin{abstract}
The Gaia–Sausage/Enceladus (GSE) structure, widely recognized as the most recent major accretion event experienced by our Galaxy, is traditionally interpreted as the remnant of a single ancient merger that played a significant role in building the Milky Way’s inner halo. Most previous studies have characterized the GSE as a kinematically coherent population that originated from either a single progenitor or a recent infall event. Here, we present evidence for a more complex origin, based on data from the  DESI and a novel unsupervised clustering algorithm, GS$^3$ Hunter. Applying this method to local halo stars near the solar neighborhood, we identify 17 structures, including known systems such as Sequoia and GSE, as well as several previously unrecognized structures/stellar streams. A more detailed analysis incorporating chronological, dynamical, and chemical dimensions reveals four distinct substructures within the GSE region, herein designated GSE‑GSH1 (12 Gyr), GSE‑GSH2 (10 Gyr), GSE‑GSH3 (8 Gyr), and GSE‑GSH4 (7 Gyr). Although all four are broadly consistent with the overall phase‑space distribution and abundance patterns of the GSE, they display markedly distinct orbital actions and chemical abundances relative to previously reported results. This finding reveals an unprecedented level of internal complexity in the GSE's formation history and supports a scenario in which the GSE is not the remnant of a single accretion event, but rather a composite structure assembled through multiple, sequential merger episodes during the early Milky Way.
\end{abstract}

The Galactic halo was assembled through a series of merger and accretion events over the course of the Milky Way’s evolution. These interactions produced numerous stellar streams and substructures in the halo, which retain distinct chemical and dynamical signatures of their progenitors\cite{2021ApJ...909L..26B,2025NewAR.10001713B,2019ApJ...880...38B}. Such signatures provide valuable insights into the sequence of merger events that shaped the Galaxy. A major outcome of this hierarchical assembly process is the so-called Ancient Last Major Merger—the Gaia–Enceladus/Sausage (GSE) event—whose remnants offer fundamental clues to the Milky Way’s early formation and accretion history\cite{2020ARA&A..58..205H, 2021ApJ...911L..21K}.

The GSE structure has traditionally been interpreted as the remnant of a single major accretion event shaping the Milky Way’s inner halo \cite{2003ApJ...585L.125B, 2018Natur.563...85H, 2018MNRAS.478..611B, 2020MNRAS.497..109F, 2021MNRAS.508.1489F}. Helmi et al.\cite{2018Natur.563...85H} suggested that this merger occurred 10–13 Gyr ago, while Belokurov et al.\cite{2018MNRAS.478..611B} placed it slightly later, at 8–11 Gyr. Both studies also considered different formation contexts for GSE, reflecting distinct analyses of its early accretion environment. Recent studies have challenged the traditional view of GSE as a single major merger. Donlon et al.\cite{2022ApJ...932L..16D, 2023ApJ...944..169D, 2024MNRAS.531.1422D} suggested that the GSE is likely not a single radial merger event that is dynamically young and did not collide with the Milky Way's protodisk at early times, as previously thought. Instead, it may have collided with the Milky Way disk within the last few Gyr, emphasizing the complexity of the GSE debris and the necessity for more refined analyses, including isochrone timescale diagnostics, as well as more detailed dynamical and chemical characterizations.

Meanwhile, the identification and characterization of stellar streams and halo substructures have progressed rapidly, with initiatives such as \textit{galstreams}\cite{2023MNRAS.520.5225M} providing comprehensive catalogues. By combining chemical and dynamical information, these structures reveal the timing and properties of past mergers. More recently, the development of the Galactic-Seismology Structures and Streams Hunter (GS$^3$ Hunter) by Wang et al.\cite{2024ApJ...974..219W} provides an effective tool for uncovering the complex formation and evolutionary pathways of the Galactic halo.

\begin{figure*}[!h]
  \centering
  \includegraphics[width=0.85\textwidth]{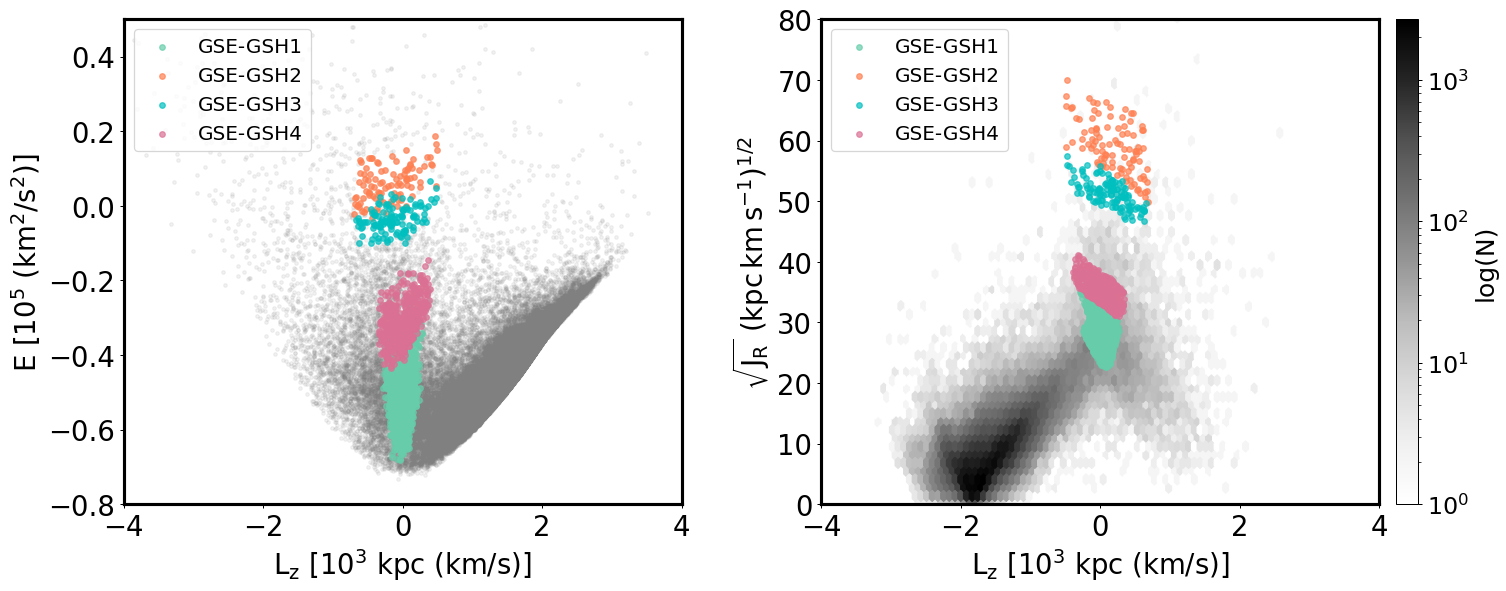}
  \caption{{\bf Dynamically Distinct Components Identified within the GSE Region.} Following the selection criteria of Zhang et al.\cite{2024ApJS..273...19Z} and Helmi et al.\cite{2018Natur.563...85H}, we identify four substructures within the GSE distribution, labeled GSE-GSH1–4. Left: distribution in angular momentum versus energy space, with each substructure highlighted in color and the full sample shown in gray. Right: distribution in $L_z$–$\sqrt{J_R}$ space, similarly color-coded. These components deviate from the main Galactic disk population, consistent with halo or accreted stars, and exhibit stratification in radial position, suggesting differences in accretion times and orbital properties. Together, these results support a multi-component, complex accretion history for the GSE.}
  \label{DESI(results)_GSE(Lz-Jr)}
\end{figure*}

% Explaination of the result in figure 1, dynamical and orbit features of the 4 components which was found in the GSE part.
We use the data from Dark Energy Spectroscopic Instrument (DESI)\cite{2016arXiv161100036D, 2024AJ....168...58D}, selecting 86,945 stars as our sample (see Section \hyperref[Data_processing_and_selection]{Data processing and selection} for details). Applying GS$^3$ Hunter to this sample reveals 17 distinct streams or substructures (see Section \hyperref[Kullback-Leibler Divergence]{Kullback-Leibler Divergence} for details). Notably, within the region associated with the GSE, we detect four separate substructures. The distribution in the $E$-$L_z$ plane (Figure~\ref{DESI(results)_GSE(Lz-Jr)}, left) reveals four distinct components, hereafter GSE-GSH1–4. Rather than a distinct separation, these sub-structures are primarily stratified by their orbital binding energy in the $E$–$L_z$ space (Figure~\ref{DESI(results)_GSE(Lz-Jr)}, left). This energy gradient and their distributions in the $\sqrt{J_R}$–$L_z$ space (Figure~\ref{DESI(results)_GSE(Lz-Jr)}, right) indicate differences in orbital properties and accretion times. All four components exhibit a broad angular momentum distribution straddling $L_z$ = 0, with a slight retrograde bias, consistent with an external origin.

\begin{figure*}[!h]
  \centering
  \includegraphics[width=0.95\textwidth]{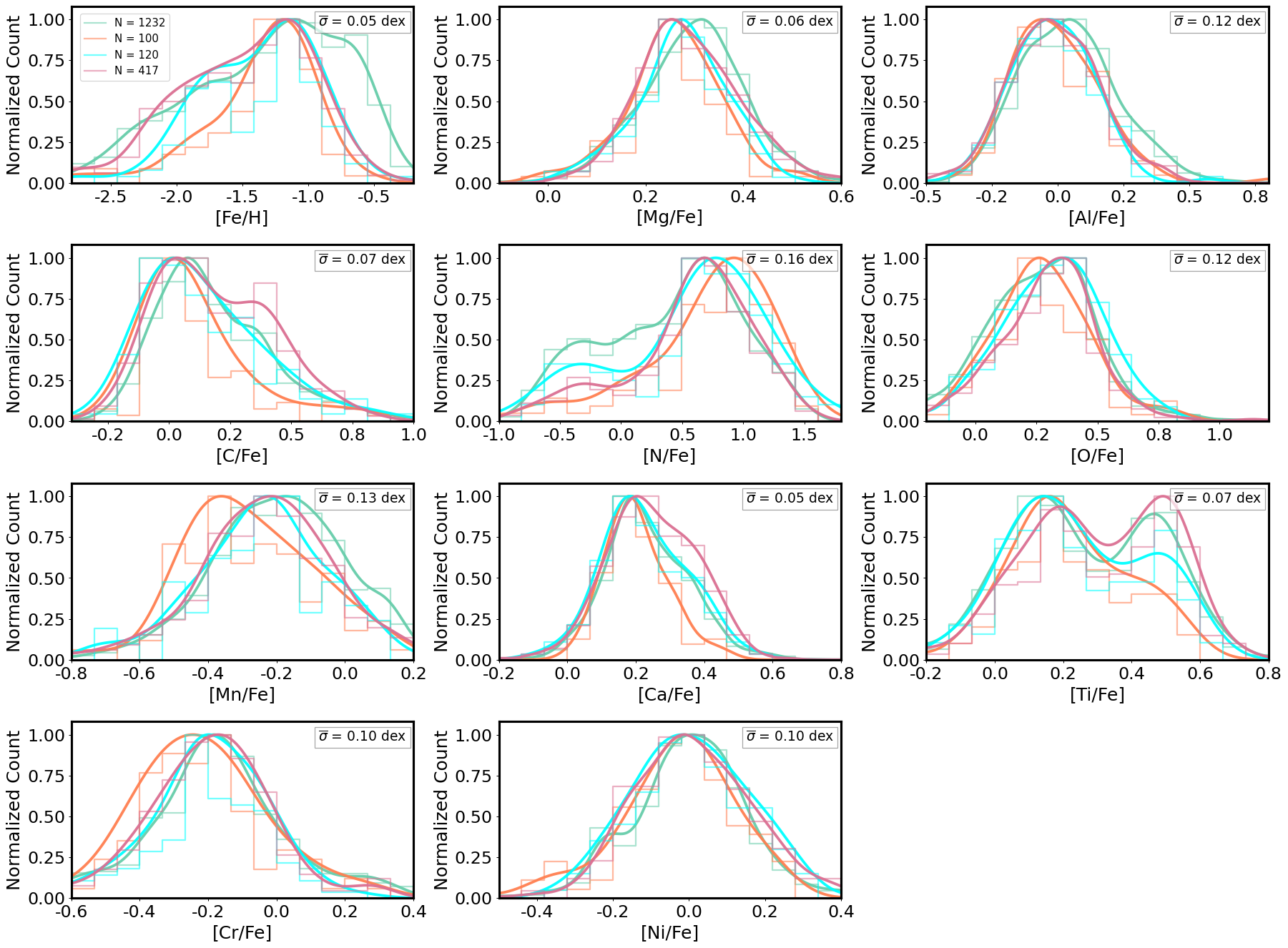}
  \caption{{\bf Chemical distribution of the different structures from GSE.} Normalized histograms and corresponding Kernel Density Estimation (KDE) curves depicting the chemical abundance distributions for the four identified GSE components. Each substructure is represented by a unique color. In the first panel, labels indicate the number of stars in each structure. Additionally, the $\sigma$ values of the corresponding chemical abundances are noted in the upper right corner of each panel.
  \label{DESI(results)_GSE(chemical abundance_hist)}}
\end{figure*}

\begin{figure*}[!h]
  \centering
  \includegraphics[width=0.95\textwidth]{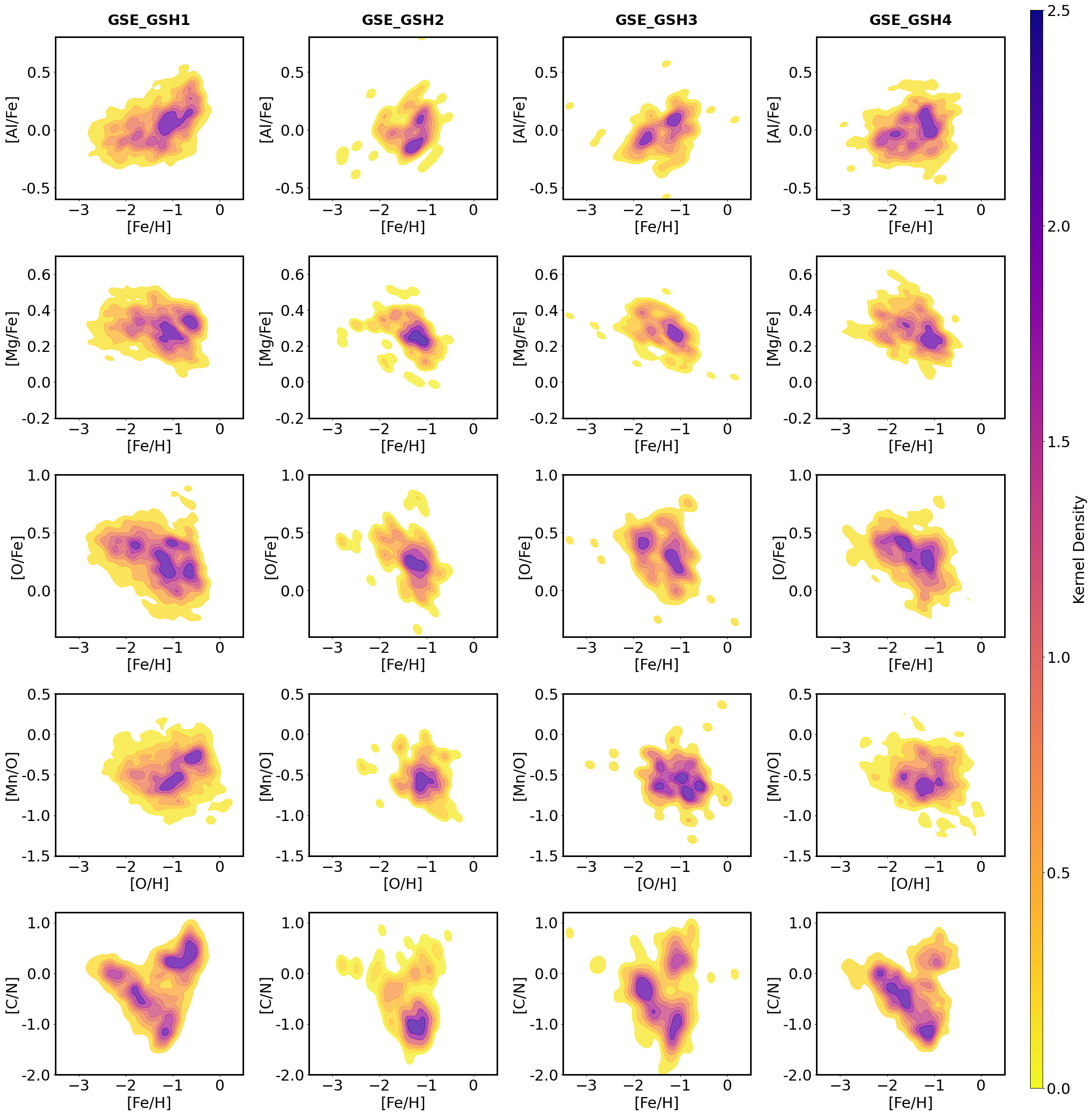}
  \caption{{\bf Chemical distribution of the GSE results.} This figure presents the KDE distributions of the four GSE-related structures in chemical abundance space. A colorbar on the right indicates the density levels. In many of the chemical abundance planes, multiple clumps are visible, which should correspond to distinct chemical evolution pathways and formation history.
  \label{DESI(results)_GSE(chemical abundance)}}
\end{figure*}

% Chemical part (Distinct Chemical Abundances and Internal Substructures), Explaination about the distribution of the 4 components in the different checmical parameters space, mainly about Figure 2 and 3. further in.
The chemical abundance distributions of the four GSE components exhibit both shared enrichment features and clear internal variations that trace their distinct accretion phases (Fig.~\ref{DESI(results)_GSE(chemical abundance_hist)}; see \hyperref[Chemical parameters of the 4 components]{Chemical parameters of the 4 components} for detailed analysis). The $\alpha$-elements — [Mg/Fe], [Ca/Fe], and [Ti/Fe] — show tight and symmetric distributions with dispersions of $\sigma \lesssim 0.07$ dex. In the [Mg/Fe]–[Fe/H] planes (Fig.~\ref{DESI(results)_GSE(chemical abundance)}), the four substructures collectively trace a high-$\alpha$ sequence at low metallicities ($[{\rm Fe/H}] \lesssim -1.5$), followed by a gradual decline toward solar [$\alpha$/Fe] as metallicity increases. The light elements [O/Fe], [C/Fe], and [N/Fe] exhibit distinct distribution patterns (Fig.~\ref{DESI(results)_GSE(chemical abundance_hist)}). The [O/Fe] distribution shows a broader spread ($\sigma \approx 0.12$ dex) and a high-abundance tail. Carbon and nitrogen display larger scatter and more complex, often asymmetric distributions. The [C/Fe] histogram presents moderate dispersion and a non-Gaussian shape, while [N/Fe] is broader and skewed, with an extended high-[N/Fe] tail in several components. The [C/N]–[Fe/H] distributions for the four components (Fig.~\ref{DESI(results)_GSE(chemical abundance)}, last row) exhibit distinct multi-lobed structures. These features indicate that these components are chemically heterogeneous, potentially comprising stellar populations with a broad range of ages. Alternatively, such structures serve as a fossil record of non-uniform mixing and the complex assembly history of the GSE accretion event.

For the light odd-Z element Al, the [Al/Fe] distribution shows a moderate spread ($\sigma \approx 0.12$ dex) across the sample, with most stars clustering between –0.4 and +0.6 dex. The main GSE population exhibits a roughly symmetric distribution peaking near solar-scaled [Al/Fe], whereas the smaller components (GSE-GSH2, GSE-GSH3, GSE-GSH4) show similar central values but slight differences in the tails. In GSE-GSH1 and GSE-GSH4, [Al/Fe] rises mildly from low metallicities (–2.5 dex) toward [Fe/H] $\approx$ –1.0 dex, followed by a plateau. By contrast, GSE-GSH2 and GSE-GSH3 display lower [Al/Fe] at a given [Fe/H], without a clear high-[Al/Fe] plateau (Fig.~\ref{DESI(results)_GSE(chemical abundance)}). For the iron-peak elements Mn, Cr, and Ni, the distributions also differ among the subcomponents. [Mn/Fe] shows a relatively large scatter ($\sigma \approx 0.13$ dex), peaking around $-$0.3 dex in the main population, with a long tail toward higher values. [Cr/Fe] displays a narrower distribution ($\sigma \approx 0.10$ dex) centred near solar. [Ni/Fe] has a similar scatter to Cr ($\sigma \approx 0.10$ dex) and a slightly subsolar mean.

\begin{figure*}[!h]
  \centering
  \includegraphics[width=0.85\textwidth]{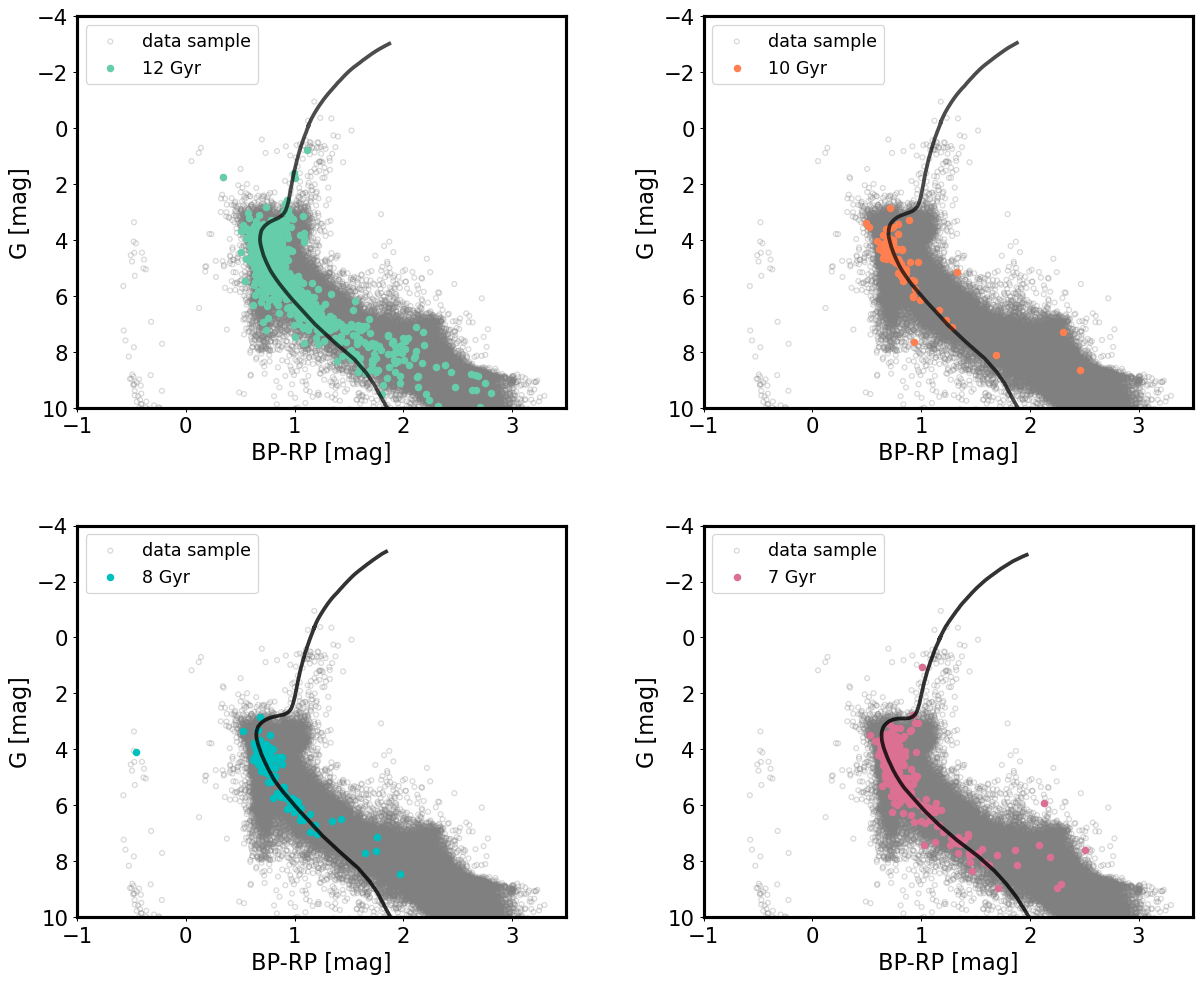}
  \caption{{\bf Age of the GSE results.} We present the CMDs of the four components, overlaid with their corresponding PARSEC isochrones \cite{2012MNRAS.427..127B} shown as solid black lines. The inferred ages for each component are labeled in the panels. Gray dots indicate the full sample.
  \label{DESI(results)_GSE_CMD}}
\end{figure*}

The kernel density maps in Fig~\ref{DESI(results)_GSE(chemical abundance)} reveal multiple, spatially separated overdensities in $\alpha$–[Fe/H] space within each substructure. These features likely trace separate star-forming regions or episodes in the progenitor systems, with each overdensity representing a stellar population formed from an interstellar medium (ISM) parcel of specific chemical composition. Their persistence indicates that the progenitors were not fully chemically homogeneous, potentially reflecting localized star formation or episodic gas accretion \cite{2001NuPhA.688..396T}. Isochrone-based age dating (Fig.~\ref{DESI(results)_GSE_CMD}) reinforces this scenario, showing components spanning $\sim$12 to $\sim$7 Gyr. Such a wide age distribution is inconsistent with a single, short-lived star formation event in a single progenitor, instead pointing to the accretion of progenitor systems with extended and diverse star formation histories.

% Comparison with Previous Work (Specialll with Haidi Newberg et al.2023), mainly about the differences about the 4 components.

Our previous analysis identified four distinct stellar substructures within the GSE region, largely consistent with those reported by Donlon et al. \cite{2023ApJ...944..169D} and independently recovered in our earlier study \cite{2024ApJ...974..219W}. In this work, however, we find notable differences among them. These components occupy partially separated loci in the $L_z$–$\sqrt{J_R}$ plane, exhibit more complex chemical abundance patterns, and span different stellar age ranges for the first time, pointing to distinct accretion stages and evolutionary histories.

Combined chemo‑dynamical and age analyses further suggest that the GSE is unlikely to be the remnant of a single, monolithic merger. Within the canonical GSE region, we identify four chemically and dynamically coherent components, each occupying distinct regions in elemental‑abundance space ([Mg/Fe], [Al/Fe], [O/Fe], [C/N], [Fe/H]) and orbital‑action space, and showing systematically different stellar ages. While all components share the high‑eccentricity, low‑angular‑momentum signature of radial infall, their offsets in action space and chemical abundances imply formation in separate star‑forming environments and accretion at different times pointing towards multiple, sequential merger episodes. The preservation of distinct chemo‑dynamical identities indicates incomplete phase mixing and supports a hierarchical, multi‑epoch assembly of the inner halo.

Recent work by Donlon et al. \cite{2024MNRAS.531.1422D} proposed that the Milky Way's "last major merger" occurred much more recently (within the last 1-2 Gyr) based on phase‑space caustics and limited phase mixing in Gaia DR3, contrasting with the traditional interpretation of GSE as an early ($\sim$8–11 Gyr) accretion event. In contrast, our results reveal a broad stellar age range spanning $\sim$7 to $\sim$12 Gyr and distinct chemo‑dynamical substructures within the canonical GSE region, the clear timescale picture of the accretion, which has never been found before, suggests a more complex assembly history that cannot be explained by a single recent merger alone. This indicates that the GSE is unlikely to be as young as proposed by Donlon et al.\cite{2024MNRAS.531.1422D}. Our current results differ from those obtained in our previous analysis based on GALAH data, possibly owing to the different sky coverage of the two surveys, with GALAH primarily sampling the southern sky and DESI covering the northern sky. Taken together, the chemical distinctions, dynamical offsets, and age spread strongly favor a brand new, multi‑event origin for the GSE.

%------------------------------------------------------------------------------------------------------------------------------------------------------------------------ 
\clearpage

\section*{References}
% \bibliographystyle{naturemag}
% \bibliography{abbrv_J,mybib2}

%%
%% TABLES
%%
%% If there are any tables, put them here.
%%

\begin{methods}
       
\subsection{Data processing and selection}
\label{Data_processing_and_selection}

This section details the data processing and analysis methods employed in this work. We use the data from the DESI\cite{2016arXiv161100036D, 2024AJ....168...58D} (Dark Energy Spectroscopic Instrument) survey. Our stellar sample is adopted from the work of Zhang et al.\cite{2024ApJS..273...19Z, 2024AJ....168...58D}, based on the DESI EDR. It consists of 520,228 sources classified as stars according to the "RR$\_$SPECTYPE`` flag. Stellar chemical abundances were derived using a data-driven Payne method\cite{2017ApJ...849L...9T, 2019ApJS..245...34X}. The uncertainties of the chemical abundances were examined for all key elements used in this work. The median uncertainties are approximately 0.03 dex for [Fe/H] and [Mg/Fe], and about 0.07 dex for [Al/Fe], while somewhat larger typical uncertainties are found for [O/Fe], [C/Fe], and [N/Fe], typically in the range of $\sim0.05$ – $0.10$ dex, with extended tails toward larger values. In addition, this work provides kinematic parameters and distance estimates for each star, enabling comprehensive chemo-dynamical analyses. The orbital energy $E$ and azimuthal and vertical actions ($J_{\phi}$, $J_{Z}$, $J_{R}$) used in this work are adopted directly from the value-added catalog of Zhang et al,\cite{2024ApJS..273...19Z} where they are computed under the MWPotential2014 Galactic potential as implemented in galpy \cite{2015ApJS..216...29B}, corresponding to a Milky Way model with a virial mass of $M_{\rm vir} \sim 0.8 \times 10^{12}\,M_\odot$. We restricted our sample to stars located within 5 kpc from the Sun. We further constrained the sample to stars with a total velocity exceeding 180 km s$^{-1}$ to select the halo stars with respect to the Local Standard of Rest (LSR) ($V_{LSR}$ = 232 km s$^{-1}$). To ensure the quality of the data sample, we removed stars with uncertainties in [Al/Fe] or [Mg/Fe] greater than 0.5 dex, we excluded stars with poorly constrained distance estimates by requiring the uncertainty in logarithmic distance to satisfy $\sigma_{\log d} < 0.5$. After applying these cuts, the distance-uncertainty distribution is strongly concentrated at low values, with most stars having $\sigma_{\log d} \lesssim 0.1$ and only a small fraction extending toward $\sim 0.3$. We excluded stars with the observing program type “other”. Additionally, stars with signal-to-noise ratio in the red arm (SNR) below 30 has been discarded. Finally, we obtain 136877 stars as our sample. To further isolate the radially anisotropic component identified in this work, we construct a dynamical subsample by selecting stars with azimuthal action $J_{\phi}$ in the range  $-700$ km s$^{-1}$ kpc $<$ $J_{\phi}$ $<$ $500$ km s$^{-1}$ kpc, eccentricity $e > 0.9$, and energy $E > -1.8 \times 10^5$ km$^{2}$ s$^{-2}$. These criteria preferentially select stars on highly radial orbits with low net rotation, consistent with expectations for Gaia–Sausage–Enceladus-like debris.

In Extended Data Fig.~\ref{data sample}, we presents the distribution of our stellar sample in the energy–angular momentum ($E$–$L_z$) plane, together with the heliocentric distance ($D_\odot$) distribution. The left and middle panels encode [Al/Fe] and [Mg/Fe] abundances, respectively, as color maps. From the [Al/Fe] panel, it is evident that the abundance of [Al/Fe] increases progressively within the range of $L_z$ from $-$4 to 4 ($\times$ 10$^3$ kpc km s$^{-1}$), with a distinct clump forming in the region where $L_z$ is greater than 0 ($\times$ 10$^3$ kpc km s$^{-1}$), with [Al/Fe] values around 0.25 dex. In contrast, the [Mg/Fe] abundance decreases steadily across the $L_z$ range from $-$4 to 4 ($\times$ 10$^3$ kpc km s$^{-1}$), with a notable clump appearing in the region where $L_z$ exceeds 1 ($\times$ 10$^3$ kpc km s$^{-1}$). The differing behaviours of [Al/Fe] and [Mg/Fe] as functions of $L_z$ may reflect variations in the star formation history (SFH) and nucleosynthetic pathways across distinct Galactic components \cite{2025arXiv251025876S}.  Stars with lower angular momentum — more likely associated with the inner halo populations — tend to exhibit higher [Mg/Fe] ratios, consistent with rapid star formation dominated by core-collapse supernovae (CCSNe). Conversely, higher $L_z$ populations, plausibly linked to the outer disk or accreted components, display distinct [Al/Fe] patterns but lower [Mg/Fe], suggesting prolonged enrichment timescales and a larger relative contribution from Type Ia supernovae.

\subsection{Disk and Splash Stars}

Aluminum and magnesium are primarily synthesized in core-collapse (Type II) supernovae, which are associated with stars that form and evolve in the disk or bulge regions of the Milky Way. The abundances of [Al/Fe] and [Mg/Fe] are sensitive to both the stellar mass of the progenitor galaxy and the star formation environment. Accreted stars, particularly those originating from low-mass dwarf galaxies, can exhibit distinct chemical abundance patterns, including, in some cases, lower [Mg/Fe] ratios at a given metallicity, consistent with their low star formation efficiency and extended chemical enrichment histories \cite{2015ApJ...799..230H, 2015MNRAS.449..761U, 2009ARA&A..47..371T}. Based on the chemical abundance plane ([Fe/H]–[Mg/Fe]), we classified the sample into three components: accreted stars, low-$\alpha$ (thin disk) stars, and high-$\alpha$ (thick disk) stars. These components are shown in the first panel of Fig.~\ref{Chem 1}. The corresponding distributions in the [Fe/H]–[Al/Fe] plane are displayed in the bottom panels. Notably, both the thick and thin disk sequences appear to exhibit substructures of two overlapping populations. This is likely a consequence of our selection criteria, which rely solely on chemical abundance parameters and thus may not fully disentangle the populations. In particular, around [Fe/H] $\approx -0.5$, the thick and thin disk loci overlap significantly, reflecting the transitional nature of stars in this metallicity regime and the limitations of purely chemical separation \cite{2019A&A...632A...4D}.

\subsection{Kullback-Leibler Divergence}
\label{Kullback-Leibler Divergence}

In this work, we use the GS$^3$ Hunter method for identifying and analyzing cluster candidates. While previous methods have made valuable contributions to identifying dynamical groups and substructures, GS$^3$ Hunter offers several advantages. By integrating both Mahalanobis distance and Euclidean distance, this approach greatly improves the accuracy of the clustering results. Furthermore, the incorporation of deep learning-based neural networks increases the efficiency of our method, making it particularly well-suited for high-dimensional, large-scale datasets. Additionally, GS$^3$ Hunter is capable of simultaneously detecting both cold and hot stellar streams, thus broadening its applicability across a range of Galactic structures. For detailed steps and procedures referred to the work by Wang et al.\cite{2024ApJ...974..219W}. After applying our method, a total of 27 cluster/group candidates were identified. 

In addition to the established procedures, we introduced a new step based on Kullback-Leibler Divergence\cite{kullback1951information} (KLD) to enhance the automation and intelligence of the results, specifically by improving the relationship between candidates and their corresponding structures. This approach was used to automatically assign the cluster/group candidates, defined through density peaks clustering algorithm\cite{2014Sci...344.1492R} (DPCA), to their corresponding structures. Defined as:

\begin{equation}
D_{\mathrm{KL}}(P \parallel Q) = \int_A P(x)\,\log\!\left(\frac{P(x)}{Q(x)}\right)\, d^2x
\end{equation}
Here, $P(x)$ describes the probability distribution jointly defined over two variables, while $Q(x) = \prod_i P_i(x_i)$ is obtained by multiplying the marginal distributions of each variable independently. The KLD was applied to quantify the similarity between the candidates, allowing us to perform a clustering process based on these similarities. Following this, an assignment has been set, resulting in the identification of 17 distinct structures. The results are shown in Extended Data Fig.~\ref{KL Divergence}, where the heatmap visualizes the KLD values among the 27 cluster candidates obtained from DPCA. In this figure, the darker the purple color, the more similar the two candidates are. Moreover, the shorter the length of the dendrogram, the greater the similarity between the candidates.

\subsection{Chemical parameters of the 4 components}
\label{Chemical parameters of the 4 components}

The small dispersions in [$\alpha$/Fe] ($\sigma \lesssim 0.07$ dex) indicate that the interstellar medium in the progenitor systems were chemically well mixed during the main phase of $\alpha$-element production (Fig.~\ref{DESI(results)_GSE(chemical abundance_hist)}). The observed high-$\alpha$ sequence at low metallicities reflects enrichment dominated by CCSNe, while the subsequent decline toward solar [$\alpha$/Fe] at higher metallicities is naturally explained by the delayed iron contribution from Type Ia supernovae, which add iron but only minor amounts of $\alpha$-elements\cite{2013ARA&A..51..457N}. The broader spread and high-[O/Fe] tail may partially reflect residual systematics in the non-local thermodynamic equilibrium (LTE)–uncorrected measurements \cite{2016MNRAS.463.1518A}, but are also consistent with inhomogeneous chemical enrichment from early SN II events, where localized feedback and turbulent mixing shape abundance patterns \cite{2012A&A...538A..82R}. Carbon and nitrogen show larger scatter due to the superposition of multiple enrichment sources. Intermediate-mass asymptotic giant branch (AGB) stars produce significant abundances of carbon on longer timescales with metallicity-dependent yields, so the [C/Fe] distribution encodes a mixture of prompt and delayed inputs and is sensitive to the progenitor’s star-formation timescale and initial mass function (IMF) \cite{2014PASA...31...30K}. Nitrogen is primarily synthesized in AGB stars, with additional contributions from rapidly rotating massive stars in some models \cite{2012A&A...537A.146E, 2024IAUS..361..259T}. Delayed release and metallicity-dependent yields naturally lead to higher [N/Fe] ratios in systems with slower enrichment or stronger AGB contributions, resulting in larger star-to-star variance and the asymmetric [N/Fe] distributions observed\cite{2025arXiv250620436K}.

The multi-lobed [C/N]–[Fe/H] patterns can be interpreted as a combination of intrinsic differences in birth [C/N] among distinct populations (reflecting different star-formation histories and AGB contributions) \cite{2015MNRAS.453.1855M} and differences due to stellar evolutionary state (mass/age) and mixing depth. Lower [C/N] at fixed [Fe/H] typically indicates more advanced dredge-up (older/evolved giants), whereas higher [C/N] points to less processed envelopes or higher natal C/N. Multiple peaks or ridges thus provide evidence for age/formation-time spreads or multiple chemically distinct star-forming zones within the progenitors \cite{2023ApJ...942...35C}.

The broader [Al/Fe] distribution compared with $\alpha$-elements such as Mg or Ca reflects aluminium’s mixed nucleosynthetic origins \cite{2011ApJ...729...16K}. The mild increase of [Al/Fe] in GSE-GSH1 and GSE-GSH4 at low metallicities, followed by a plateau, is consistent with early enrichment dominated by CCSNe. The lower [Al/Fe] and absence of a high-[Al/Fe] plateau in GSE-GSH2 and GSE-GSH3 may indicate slower chemical evolution or reduced contribution from massive stars prior to the onset of Type Ia supernovae \cite{2016A&A...589A.115S}.

The iron-peak elements show patterns reflecting their nucleosynthetic origins and the chemical evolution of the GSE substructures. The subsolar [Mn/Fe] in the main population is consistent with early enrichment by metal-poor CCSNe, whose Mn yields depend on metallicity\cite{2013ARA&A..51..457N}, while the high-[Mn/Fe] tail likely reflects later contributions from Type Ia supernovae\cite{2025arXiv250620436K}. [Cr/Fe], with a narrow distribution around solar, suggests co-production with Fe in both core-collapse and thermonuclear supernovae with weak metallicity dependence. [Ni/Fe] also traces both channels, though subtle variations may indicate differences in progenitor mass distributions or neutron excess in the explosive burning zones \cite{2024A&A...682A.116N}.

The two-dimensional [X/Fe]–[C/N] kernel-density maps (Extended Data Fig.~\ref{DESI(results)_GSE_CN(age)}) for the four GSE substructures display several key features that shed light on the chemical evolution of these components. The distributions of [Al/Fe], [Mg/Fe], and [O/Fe] exhibit well-defined, compact regions in the [C/N]–[Fe/H] space, with a clear correlation between [C/N] and [X/Fe]. These distributions suggest that the progenitors of these substructures followed a relatively homogeneous enrichment pathway with a dominant contribution from CCSNe, resulting in tight [X/Fe] ratios consistent with rapid early star formation. The distinct peaks and coherent clustering of data points in each substructure indicate that local variations in SFH and chemical enrichment processes likely led to small but measurable differences in their [C/N] ratios. Notably, GSE-GSH2 departs from this general trend, showing multiple density peaks across [C/N], particularly in [Al/Fe] and [O/Fe]. This feature points to a more extended star-formation history with intermittent enrichment episodes and inefficient metal mixing, in contrast to the more homogeneous patterns observed in the other three substructures. The clear separation of these clusters suggests that even small changes in the progenitors’ SFH or IMF may leave an imprint in the chemical signature, particularly in light elements such as carbon and nitrogen \cite{2025arXiv250506606E}. Together, these differences in [C/N] distributions reinforce the view that the GSE substructures trace multiple accretion events, each preserving a distinct chemical fingerprint of its progenitor system.

The [Al/Fe]–[Mg/Fe] maps color-coded by Energy for the four GSE substructures (Extended Data Fig.~\ref{DESI(results)_GSE_chemical_E}) reveal a strong coupling between detailed chemical abundance ratios and dynamical properties, providing a direct link between the chemo-dynamical evolution of the progenitors. In all cases, coherent gradients in $E$ are present across the [Al/Fe]–[Mg/Fe] plane, indicating that stars with different orbital binding energies occupy systematically distinct loci in chemical space. Such gradients suggest that the progenitor systems were not chemically well-mixed at the time of disruption, and that stars with different chemical signatures were preferentially stripped at different phases of the accretion process.

\subsection{Orbit of the 4 components}
To further characterize the orbital structure of the four chemically identified GSE components, we examine their distribution in the action space, presented here in a polar coordinate projection (Extended Data Fig.~\ref{DESI(results)_GSE(orbit-dynamic)}). This representation emphasizes the orbital geometry, with purely circular orbits lying along the horizontal axis ($J_R$ = 0) and purely radial orbits aligned with the vertical axis ($L_z$ = 0), while prograde and retrograde motions occupy the left and right quadrants, respectively. The bulk of the general halo population occupies a broad swath of this space, with a pronounced concentration towards the radial regime, reflecting the well-known high-eccentricity nature of GSE debris. Although their clustering in the low-$L_z$, high-$J_R$ region ties them clearly to the GSE progenitor, their lack of perfect coincidence in the IOM space (Fig.~\ref{DESI(results)_GSE(Lz-Jr)}) points to a complex assembly history than a single, monolithic event.

\subsection{Age of the 4 components}
Figure~\ref{DESI(results)_GSE_CMD} shows $Gaia$ color–magnitude diagrams (CMDs) for the four components identified within the GSE region, each overlaid with the best-fitting PARSEC isochrones \cite{2012MNRAS.427..127B}. The derived ages span from $\log t = 10.08$ dex ($\sim$ 12 Gyr) to $\log t = 9.85$ dex ($\sim$ 7 Gyr), revealing significant differences among the stellar populations. The presence of such a wide age spread, together with the chemical and kinematic diversity described in previous sections, strongly argues against the GSE being the remnant of a single merger event. Instead, these findings point to GSE as the composite outcome of multiple accretion episodes, potentially involving progenitors with different star formation histories, and enrichment timescales.

The chemical abundance patterns ([Mg/Fe], [Al/Fe], [Fe/H]) demonstrate that the four main identified components within the GSE region occupy distinct loci in multiple elemental planes, indicating divergent enrichment histories and star formation timescales. Dynamically, their distributions in action space (Extended Data Fig.~\ref{DESI(results)_GSE(orbit-dynamic)}) reveal distinct, offset loci for each component. These offsets point to subtle but significant differences in orbital eccentricity and angular momentum, consistent with material being stripped at different phases of the progenitor’s disruption, or from multiple progenitors.

The age dating from isochrone fitting (Fig.~\ref{DESI(results)_GSE_CMD}) reinforces this picture: the components span a wide range of stellar ages, from $\sim$ 12 Gyr to $\sim$ 7 Gyr. Such a spread is inconsistent with a single, short-lived star formation episode in a single progenitor, but is naturally explained by multiple, sequential accretion events. Taken together, the chemical distinctions, dynamical offsets, and age spread strongly favor a multi-event origin for the GSE, in which several accretion episodes—possibly involving progenitors of different masses and chemical evolution pathways—collectively produced the observed structure in the Milky Way’s inner halo.

\subsection{Results of the new structures identified}
After using GS$^{3}$ Hunter algorithm, we identify a total of 17 structures in our sample. By checking with previous reported structures, we recover two known structures: GSE and Sequoia. The identification of Sequoia is based on a comparison of the chemical abundance and action-space distributions of our candidate with those defined for Sequoia in Myeong et al. \cite{2019MNRAS.488.1235M}. Within the GSE, we further resolve four chemically and dynamically distinct components, shown in the first row of Extended Data Fig.~\ref{DESI(results)_LE}. Sequoia is presented in the first panel of the second row. The remaining panels in Extended Data Fig.~\ref{DESI(results)_LE} display the $L_z$–$E$ distributions of the newly discovered substructures identified in this work. We further examine the chemical properties of Sequoia and these new structures, as shown in Extended Data Fig.~\ref{DESI(results)_(chemical abundance_hist)}.

Sequoia exhibits a relatively wide metallicity range, centred at [Fe/H] $\approx -1.68$ dex with a dispersion of $\sigma \approx 0.05$ dex, consistent with previous studies and indicative of a chemically coherent progenitor. In contrast, the newly identified streams display a broader diversity in their mean metallicities, spanning from $-1.13$ to $-1.95$ dex, and in some cases showing asymmetric or multi-peaked distributions. Such diversity suggests that these streams likely originated from multiple progenitors with differing stellar masses and star formation histories. The relatively small internal dispersions ($\sigma \approx$ 0.04 $–$ 0.06 dex) across most streams point to chemically homogeneous parent systems.

% Among the 27 cluster members, two are excluded from the final visualization due to low star counts. The KLD-based merging yielded 17 distinct kinematic structures, which are shown in Figure \ref{KL Divergence}. Notably, this includes the Gaia–Sausage/Enceladus (GSE) and Sequoia, two well-established merger remnants. The remaining 13 structures are previously unreported and exhibit diverse phase-space properties.

% Figure \ref{DESI(results)_LE} displays these 17 structures in the energy–angular momentum (E–L$_z$) plane, highlighting their kinematic separability. Complementarily, Figure \ref{DESI(results)_(chemical abundance_hist)} presents the [Fe/H] distributions of all structures, including measurement uncertainties, revealing variations in their chemical properties.

\end{methods}

\begin{addendum}
 \item  G.Y.W. gratefully acknowledges financial support from from the China Scholarship Council (CSC). This work has made use of the data from the DESI Member Institutions (\url{https://www.desi.lbl.gov/collaborating-institutions}), processed by Data-driven payne method (DD-PAYNE). The Guo Shou Jing Telescope (the Large Sky Area Multi-Object Firber Spectroscopic Telescope, LAMOST) is a National Major Scientific Project built by the Chinese Academy of Sciences. Funding for the project has been provided by the National Development and Reform Commission. LAMOST is operated and managed by National Astronomical Observatories, Chinese Academy of Sciences. This work has also made use of data from the European Space Agency (ESA) mission {\it Gaia} (\url{https://www.cosmos.esa.int/gaia}), processed by the {\it Gaia} Data Processing and Analysis Consortium (DPAC, \url{https://www.cosmos.esa.int/web/gaia/dpac/consortium}). Funding for the DPAC has been provided by national institutions, in particular the institutions participating in the {\it Gaia} Multilateral Agreement. 
 
 \item[Authors contributions] H.F.W. conceived the initial idea, developed the sample selection strategy and the overall logical framework of the manuscript, and contributed to the writing. G.Y.W. developed and tested the code, performed the sample selection, and contributed to the writing alongside the other co‑authors. G.C., G.T., J.B., and T.T. contributed to the discussion of the results. All authors reviewed and helped improve the manuscript.

 \item[Competing Interests] The authors declare that they have no competing financial interests.
 
%  \item[Correspondence and request for materials] should be addressed to Eloisa Poggio~(email: eloisa.poggio@inaf.it). 
%  \item[Data availability] The dataset can be downloaded at \url{https://figshare.com/articles/Giants_P18_csv/11382705}.
\end{addendum}

\section*{Data Availability}
The data that support the findings of this study are available in the National Astronomical Data Center (NADC) at \url{https://nadc.china-vo.org/res/r101471/}.

\section*{Code Availability}
The custom code used for data analysis and generating the figures in this study is available from the corresponding author upon reasonable request.

\clearpage
\section*{Extended Data}

\setcounter{figure}{0} 
\renewcommand{\figurename}{Extended Data Figure} 
\renewcommand{\theHfigure}{ED.\arabic{figure}}

\begin{figure*}[h]
  \centering
  \includegraphics[width=\textwidth]{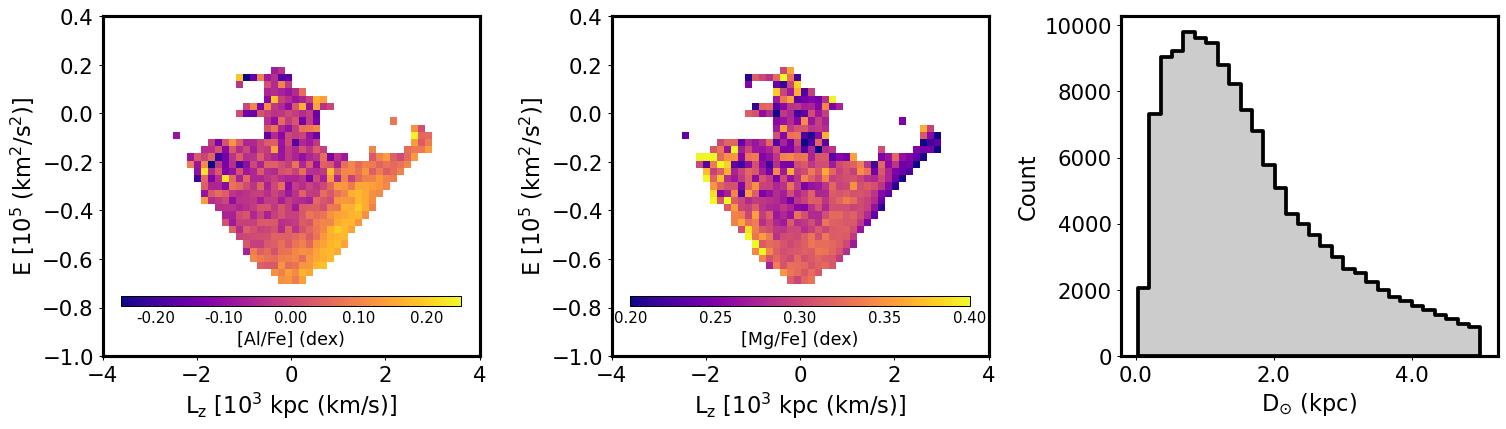}
  \caption{{\bf Distribution of the DESI sample.} This figure illustrates the spatial and chemo-dynamical distribution of the stellar sample. The left and middle panels show the sample in energy–angular momentum ($E$–$L_z$) space, color-coded by [Al/Fe] and [Mg/Fe] abundance ratios, respectively. The right panel presents the histogram of heliocentric distances ($D_\odot$).
  \label{data sample}}
\end{figure*}

\begin{figure*}
  \centering
  \includegraphics[width=\textwidth]{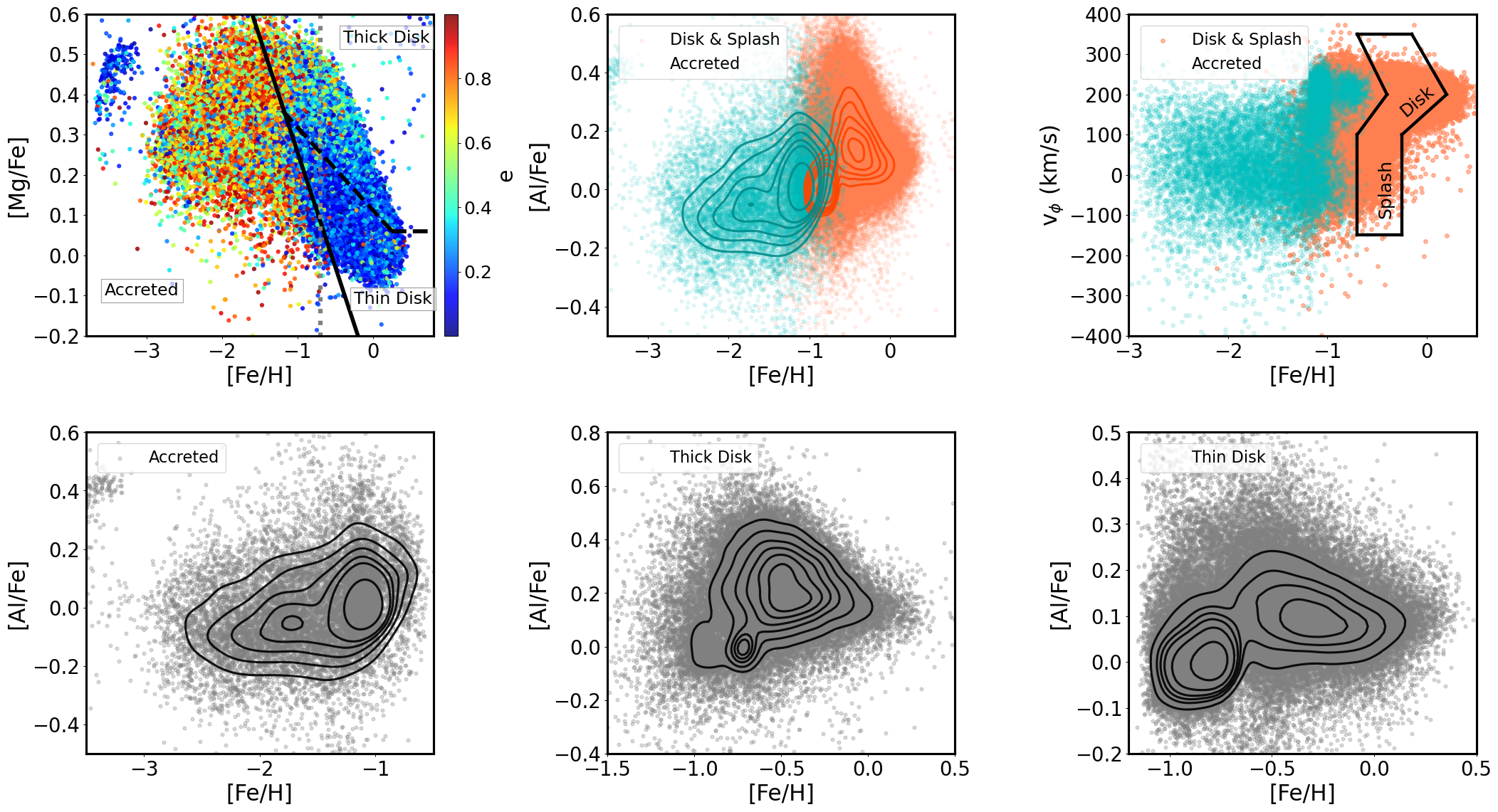}
  \caption{{\bf Chemical distribution of the DESI sample.} We separate the accreted, thick-disk, and thin-disk populations using the selection boundaries adopted from Mackereth et al. \cite{2019MNRAS.482.3426M}, as shown in the first panel. The corresponding stellar distributions for these three components are displayed in the second row. In the subsequent panels, the accreted and disk populations are highlighted in cyan and orange, respectively, in both the [Al/Fe]–[Fe/H] and $v_\phi$–[Fe/H] planes. The disk and “Splash” components are further selected following the criteria of Belokurov et al. \cite{2020MNRAS.494.3880B}.
  \label{Chem 1}}
\end{figure*}

\begin{figure*}
  \centering
  \includegraphics[width=0.8\textwidth]{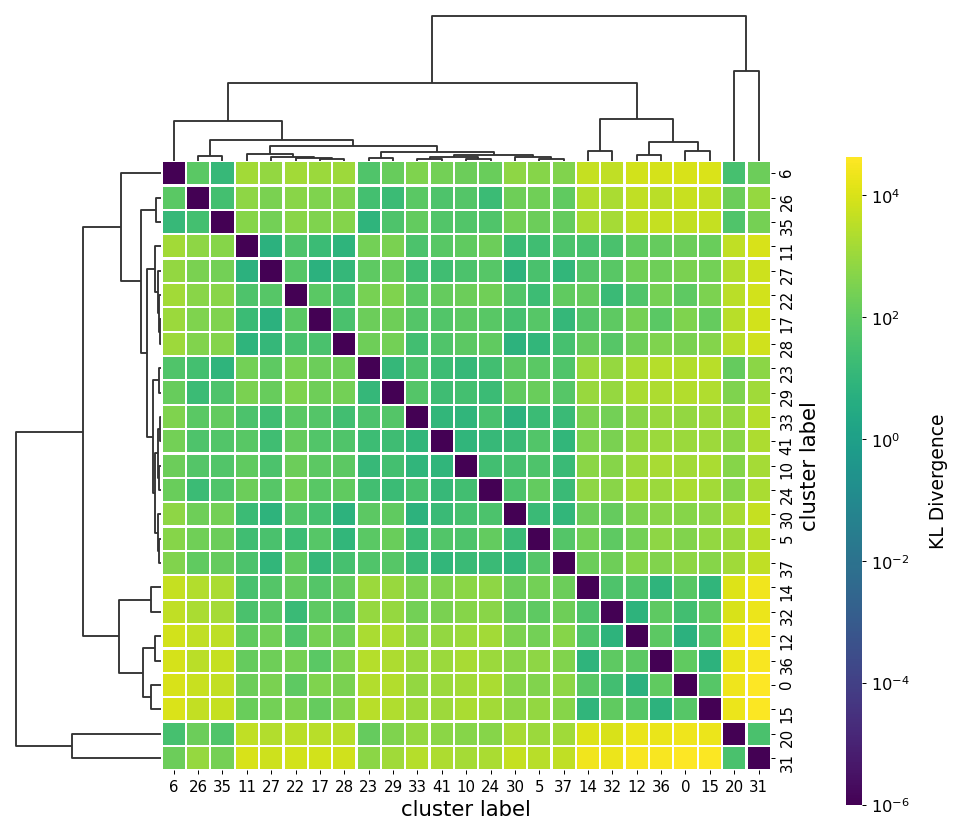}
  \caption{{\bf KL Divergence of the GS Hunter results.} This figure presents the heatmap and dendrogram of the similarity among the 25 cluster members derived using Kullback–Leibler divergence (KLD). The divergence values are indicated by the colorbar, with smaller KL values representing higher similarity between clusters and appearing in deeper purple shades. The accompanying dendrogram also illustrates the cluster members similarity: the shorter the linkage between two clusters, the more similar they are. Note two additional clusters with very small membership are not shown here, as their limited sample sizes lead to unstable and uninformative KL divergence estimates.} 
  \label{KL Divergence}
\end{figure*}

\begin{figure*}[h]
  \centering
  \includegraphics[width=\textwidth]{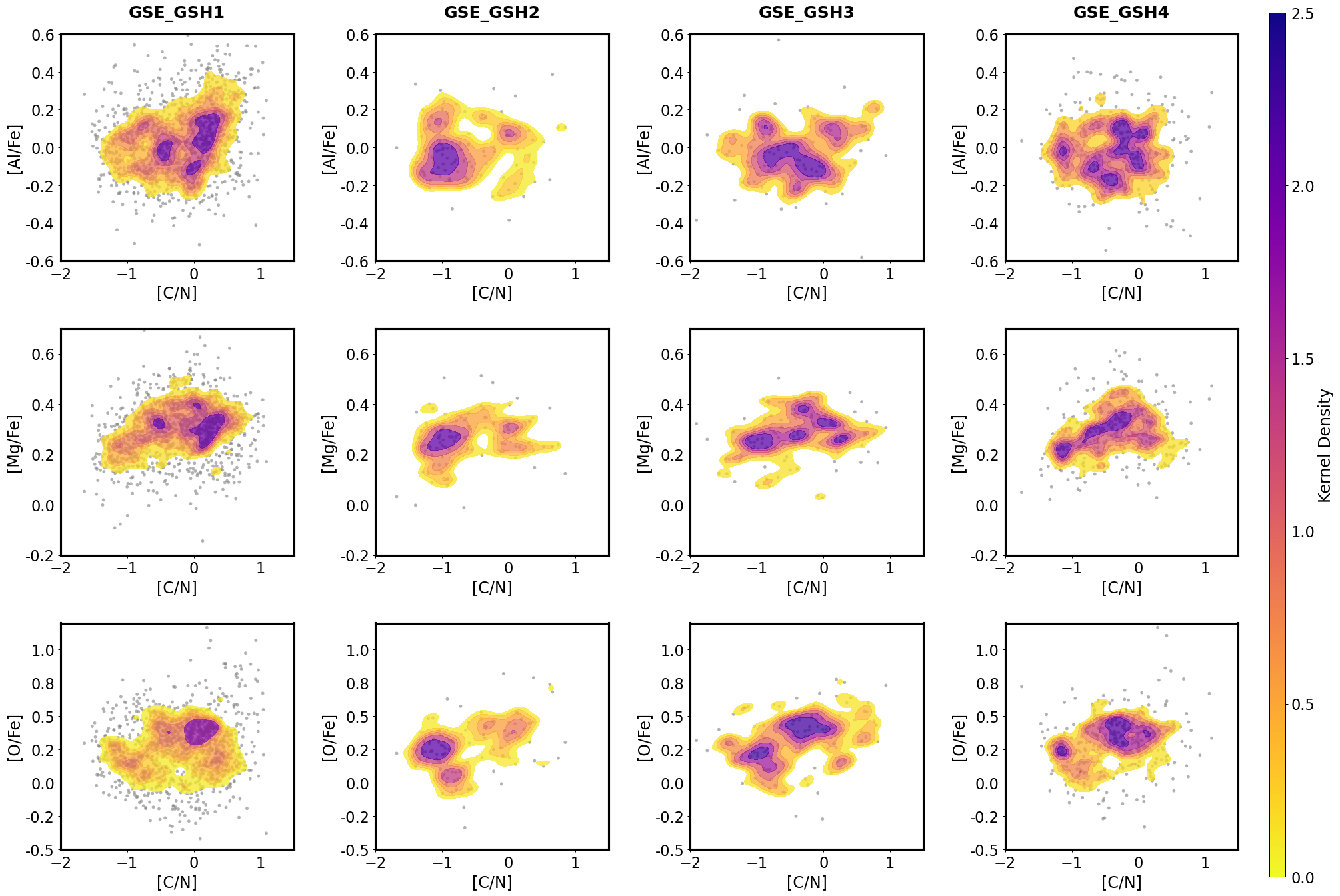}
  \caption{{\bf [C/N] vs. [X/Fe]  distribution of the GSE results.} This figure shows the KDE distributions of the GSE results in the [C/N] vs. [X/Fe] chemical space, with density levels indicated by the colorbar on the right. The [C/N] suggest a wide distribution in stellar ages. Since [C/N] is known to decrease as stars evolve along the red giant branch due to internal mixing, higher [C/N] values correspond to lower-mass progenitors that evolve more slowly, thus representing older stellar populations, while lower [C/N] values indicate higher-mass, relatively younger stars. Multiple clumps are clearly visible, possibly reflecting a mixture of stellar populations formed through different star formation episodes or accretion events associated with GSE.
  \label{DESI(results)_GSE_CN(age)}}
\end{figure*}

\begin{figure*}[h]
  \centering
  \includegraphics[width=0.85\textwidth]{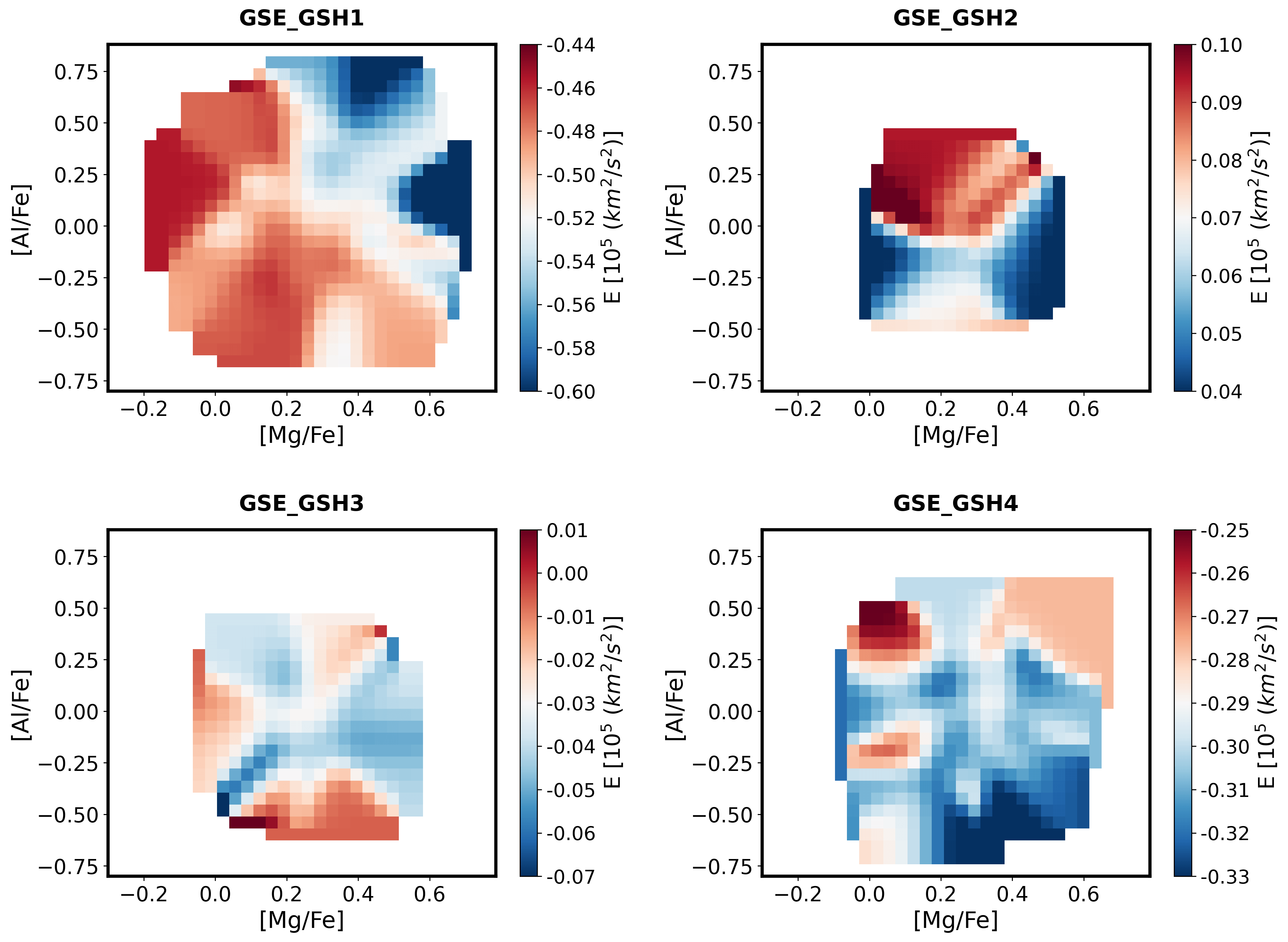}
  \caption{{\bf Chemical vs. Energy distribution of the GSE results.} This figure presents the chemical abundance distributions of the four GSE structures, with energy represented by a colorbar. A clear gradient in energy is observed across the chemical space.
  \label{DESI(results)_GSE_chemical_E}}
\end{figure*}

\begin{figure*}[h]
  \centering
  \includegraphics[width=0.75\textwidth]{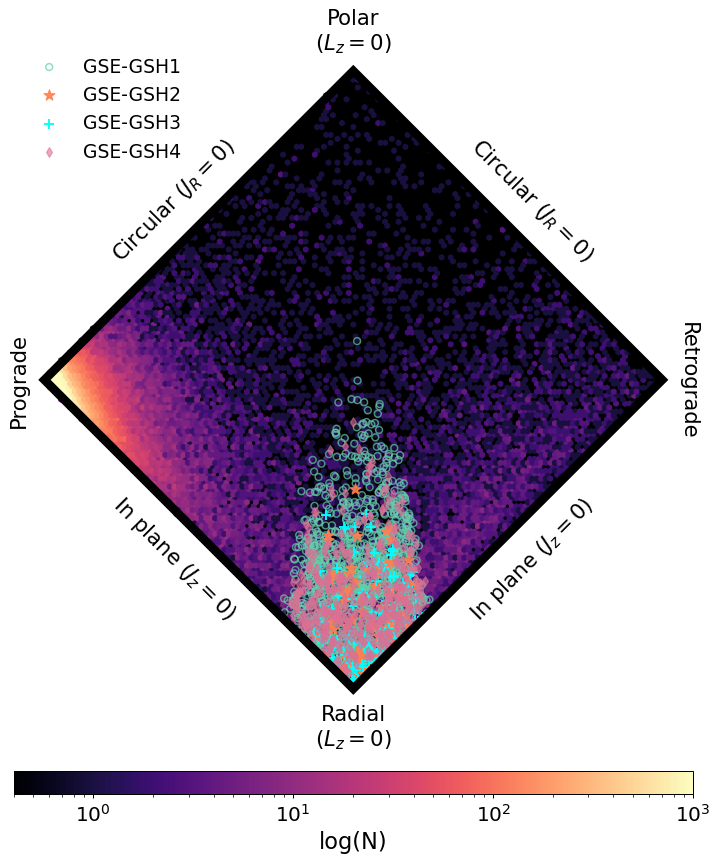}
  \caption{{\bf Orbit distribution of the GSE results.} This figure shows the distribution of the four GSE-identified structures in action space, with each structure highlighted in different colors. The full sample of stars is also displayed, with their density indicated by a colorbar. These four structures are characterized by relatively low $L_z$ and high $J_R$. However, their distributions appear significantly overlapping.
  \label{DESI(results)_GSE(orbit-dynamic)}}
\end{figure*}

\begin{figure*}[h]
  \centering
  \includegraphics[width=\textwidth]{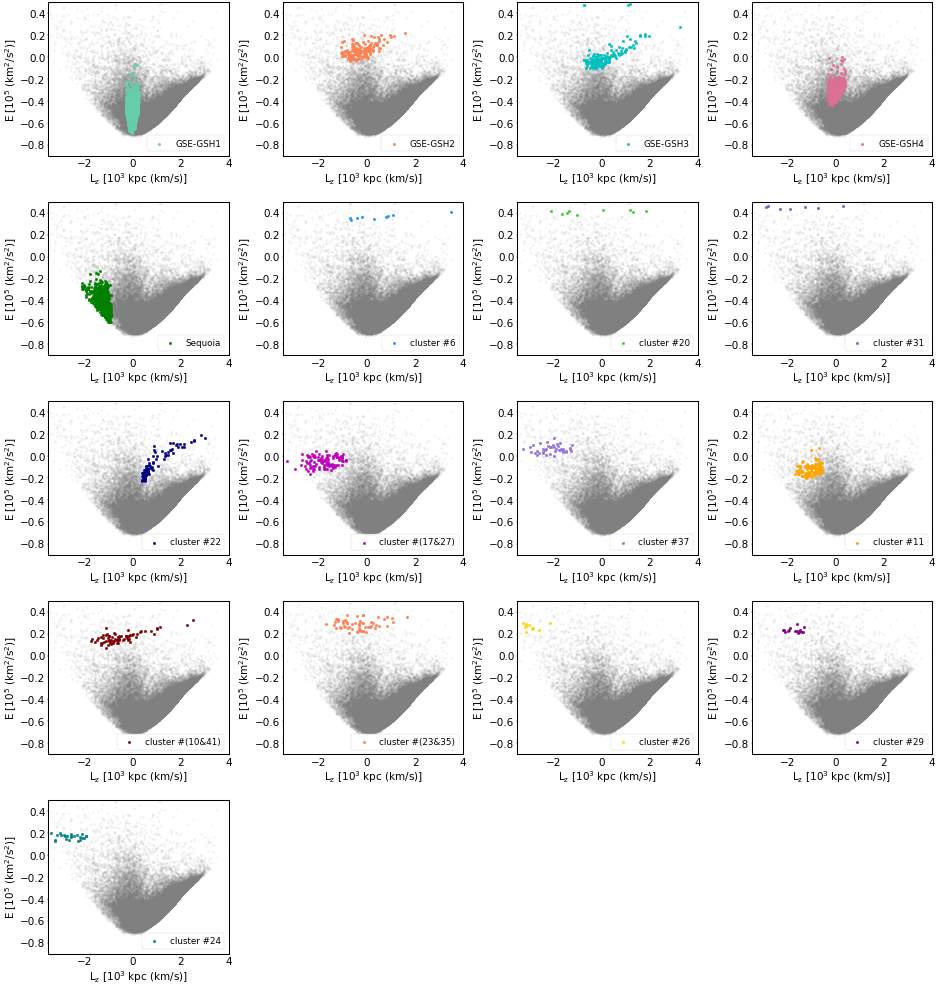}
  \caption{{\bf Results in the L$_z$--E distribution.} This figure shows the distribution of the 17 structures identified through KLD in the $L_z$–$E$ space. Among them, we recover known structures such as GSE and Sequoia, along with several unidentified substructures, each labeled in its respective panel. Gray dots indicate the full sample.
  \label{DESI(results)_LE}}
\end{figure*}

\begin{figure*}[h]
  \centering
  \includegraphics[width=\textwidth]{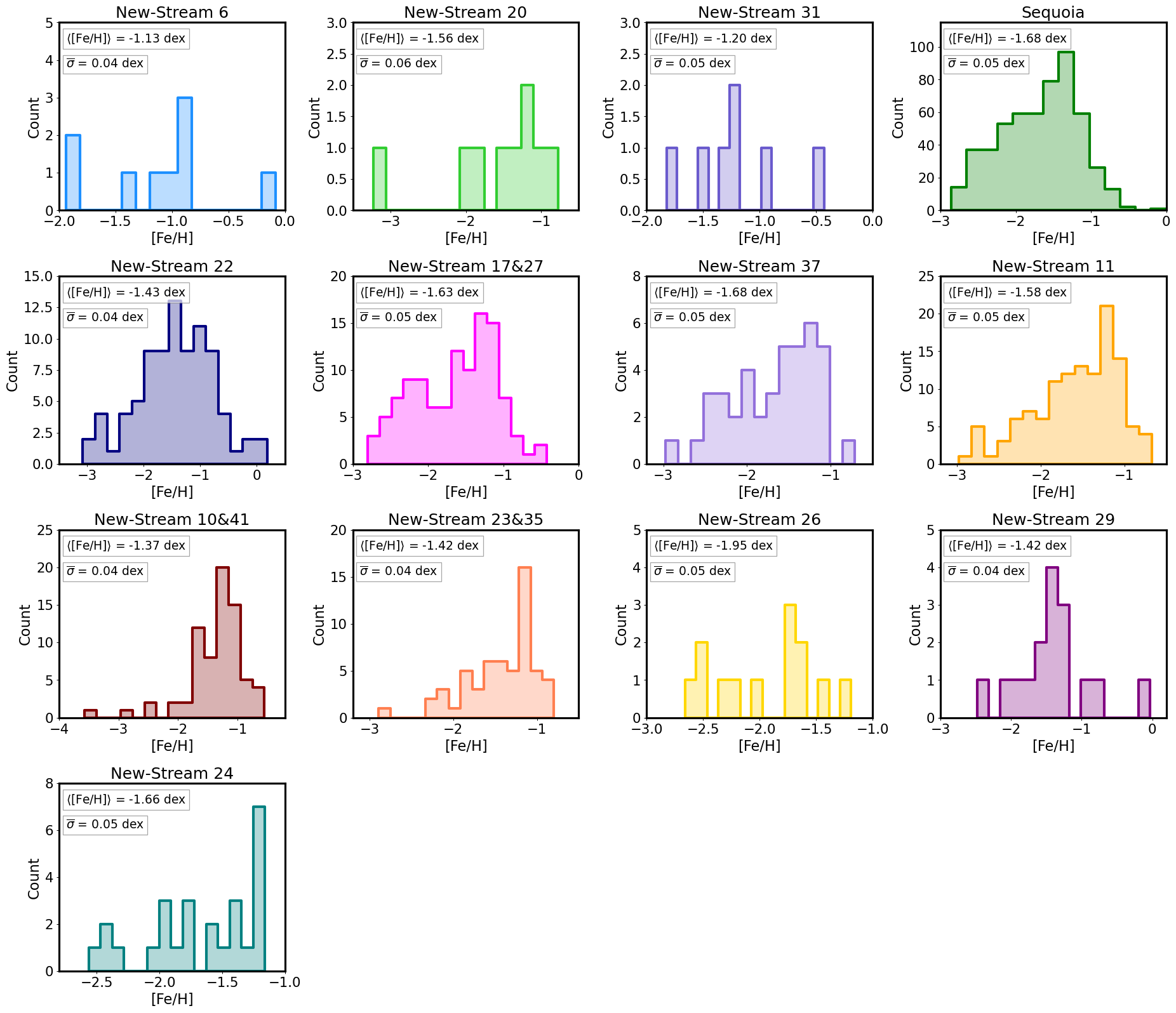}
  \caption{{\bf Results in the chemical distribution.} We present the histograms for the Sequoia and the newly identified structures, with the corresponding [Fe/H] values and standard deviations ($\sigma$) indicated in the upper-left corner of each panel.
  \label{DESI(results)_(chemical abundance_hist)}}
\end{figure*}


\begin{thebibliography}{10}
\expandafter\ifx\csname url\endcsname\relax
  \def\url#1{\texttt{#1}}\fi
\expandafter\ifx\csname urlprefix\endcsname\relax\def\urlprefix{URL }\fi
\providecommand{\bibinfo}[2]{#2}
\providecommand{\eprint}[2][]{\url{#2}}

\bibitem{2021ApJ...909L..26B}
\bibinfo{author}{{Bonaca}, A.} \emph{et~al.}
\newblock \bibinfo{title}{{Orbital Clustering Identifies the Origins of Galactic Stellar Streams}}.
\newblock \emph{\bibinfo{journal}{Astrophys.\ J.\ Lett.}} \textbf{\bibinfo{volume}{909}}, \bibinfo{pages}{L26} (\bibinfo{year}{2021}).

\bibitem{2025NewAR.10001713B}
\bibinfo{author}{{Bonaca}, A.} \& \bibinfo{author}{{Price-Whelan}, A.~M.}
\newblock \bibinfo{title}{{Stellar streams in the Gaia era}} \textbf{\bibinfo{volume}{100}}, \bibinfo{pages}{101713} (\bibinfo{year}{2025}).

\bibitem{2019ApJ...880...38B}
\bibinfo{author}{{Bonaca}, A.}, \bibinfo{author}{{Hogg}, D.~W.}, \bibinfo{author}{{Price-Whelan}, A.~M.} \& \bibinfo{author}{{Conroy}, C.}
\newblock \bibinfo{title}{{The Spur and the Gap in GD-1: Dynamical Evidence for a Dark Substructure in the Milky Way Halo}}.
\newblock \emph{\bibinfo{journal}{Astrophys.\ J.}} \textbf{\bibinfo{volume}{880}}, \bibinfo{pages}{38} (\bibinfo{year}{2019}).

\bibitem{2020ARA&A..58..205H}
\bibinfo{author}{{Helmi}, A.}
\newblock \bibinfo{title}{{Streams, Substructures, and the Early History of the Milky Way}}.
\newblock \emph{\bibinfo{journal}{Annual Review of Astron. Astrophys.}} \textbf{\bibinfo{volume}{58}}, \bibinfo{pages}{205--256} (\bibinfo{year}{2020}).

\bibitem{2021ApJ...911L..21K}
\bibinfo{author}{{Kim}, Y.~K.}, \bibinfo{author}{{Lee}, Y.~S.}, \bibinfo{author}{{Beers}, T.~C.} \& \bibinfo{author}{{Koo}, J.-R.}
\newblock \bibinfo{title}{{Evidence for Multiple Accretion Events in the Gaia-Sausage/Enceladus Structures}}.
\newblock \emph{\bibinfo{journal}{Astrophys.\ J.\ Lett.}} \textbf{\bibinfo{volume}{911}}, \bibinfo{pages}{L21} (\bibinfo{year}{2021}).

\bibitem{2003ApJ...585L.125B}
\bibinfo{author}{{Brook}, C.~B.}, \bibinfo{author}{{Kawata}, D.}, \bibinfo{author}{{Gibson}, B.~K.} \& \bibinfo{author}{{Flynn}, C.}
\newblock \bibinfo{title}{{Galactic Halo Stars in Phase Space: A Hint of Satellite Accretion?}}
\newblock \emph{\bibinfo{journal}{Astrophys.\ J.\ Lett.}} \textbf{\bibinfo{volume}{585}}, \bibinfo{pages}{L125--L129} (\bibinfo{year}{2003}).

\bibitem{2018Natur.563...85H}
\bibinfo{author}{{Helmi}, A.} \emph{et~al.}
\newblock \bibinfo{title}{{The merger that led to the formation of the Milky Way's inner stellar halo and thick disk}}.
\newblock \emph{\bibinfo{journal}{Nature}} \textbf{\bibinfo{volume}{563}}, \bibinfo{pages}{85--88} (\bibinfo{year}{2018}).

\bibitem{2018MNRAS.478..611B}
\bibinfo{author}{{Belokurov}, V.}, \bibinfo{author}{{Erkal}, D.}, \bibinfo{author}{{Evans}, N.~W.}, \bibinfo{author}{{Koposov}, S.~E.} \& \bibinfo{author}{{Deason}, A.~J.}
\newblock \bibinfo{title}{{Co-formation of the disc and the stellar halo}}.
\newblock \emph{\bibinfo{journal}{Mon.\ Not.\ R.\ Astron.\ Soc.}} \textbf{\bibinfo{volume}{478}}, \bibinfo{pages}{611--619} (\bibinfo{year}{2018}).

\bibitem{2020MNRAS.497..109F}
\bibinfo{author}{{Feuillet}, D.~K.}, \bibinfo{author}{{Feltzing}, S.}, \bibinfo{author}{{Sahlholdt}, C.~L.} \& \bibinfo{author}{{Casagrande}, L.}
\newblock \bibinfo{title}{{The SkyMapper-Gaia RVS view of the Gaia-Enceladus-Sausage - an investigation of the metallicity and mass of the Milky Way's last major merger}}.
\newblock \emph{\bibinfo{journal}{Mon.\ Not.\ R.\ Astron.\ Soc.}} \textbf{\bibinfo{volume}{497}}, \bibinfo{pages}{109--124} (\bibinfo{year}{2020}).

\bibitem{2021MNRAS.508.1489F}
\bibinfo{author}{{Feuillet}, D.~K.}, \bibinfo{author}{{Sahlholdt}, C.~L.}, \bibinfo{author}{{Feltzing}, S.} \& \bibinfo{author}{{Casagrande}, L.}
\newblock \bibinfo{title}{{Selecting accreted populations: metallicity, elemental abundances, and ages of the Gaia-Sausage-Enceladus and Sequoia populations}}.
\newblock \emph{\bibinfo{journal}{Mon.\ Not.\ R.\ Astron.\ Soc.}} \textbf{\bibinfo{volume}{508}}, \bibinfo{pages}{1489--1508} (\bibinfo{year}{2021}).

\bibitem{2022ApJ...932L..16D}
\bibinfo{author}{{Donlon}, T., II}, \bibinfo{author}{{Newberg}, H.~J.}, \bibinfo{author}{{Kim}, B.} \& \bibinfo{author}{{L{\'e}pine}, S.}
\newblock \bibinfo{title}{{The Local Stellar Halo is Not Dominated by a Single Radial Merger Event}}.
\newblock \emph{\bibinfo{journal}{Astrophys.\ J.\ Lett.}} \textbf{\bibinfo{volume}{932}}, \bibinfo{pages}{L16} (\bibinfo{year}{2022}).

\bibitem{2023ApJ...944..169D}
\bibinfo{author}{{Donlon}, T.} \& \bibinfo{author}{{Newberg}, H.~J.}
\newblock \bibinfo{title}{{A Swing of the Pendulum: The Chemodynamics of the Local Stellar Halo Indicate Contributions from Several Radial Merger Events}}.
\newblock \emph{\bibinfo{journal}{Astrophys.\ J.}} \textbf{\bibinfo{volume}{944}}, \bibinfo{pages}{169} (\bibinfo{year}{2023}).

\bibitem{2024MNRAS.531.1422D}
\bibinfo{author}{{Donlon}, T.} \emph{et~al.}
\newblock \bibinfo{title}{{The debris of the 'last major merger' is dynamically young}}.
\newblock \emph{\bibinfo{journal}{Mon.\ Not.\ R.\ Astron.\ Soc.}} \textbf{\bibinfo{volume}{531}}, \bibinfo{pages}{1422--1439} (\bibinfo{year}{2024}).

\bibitem{2023MNRAS.520.5225M}
\bibinfo{author}{{Mateu}, C.}
\newblock \bibinfo{title}{{galstreams: A library of Milky Way stellar stream footprints and tracks}}.
\newblock \emph{\bibinfo{journal}{Mon.\ Not.\ R.\ Astron.\ Soc.}} \textbf{\bibinfo{volume}{520}}, \bibinfo{pages}{5225--5258} (\bibinfo{year}{2023}).

\bibitem{2024ApJ...974..219W}
\bibinfo{author}{{Wang}, G.-Y.} \emph{et~al.}
\newblock \bibinfo{title}{{Galactic-Seismology Substructures and Streams Hunter with LAMOST and Gaia. I. Methodology and Local Halo Results}}.
\newblock \emph{\bibinfo{journal}{Astrophys.\ J.}} \textbf{\bibinfo{volume}{974}}, \bibinfo{pages}{219} (\bibinfo{year}{2024}).

\bibitem{2024ApJS..273...19Z}
\bibinfo{author}{{Zhang}, M.} \emph{et~al.}
\newblock \bibinfo{title}{{Determining Stellar Elemental Abundances from DESI Spectra with the Data-driven Payne}}.
\newblock \emph{\bibinfo{journal}{Astrophys.\ J.\ Supplement}} \textbf{\bibinfo{volume}{273}}, \bibinfo{pages}{19} (\bibinfo{year}{2024}).

\bibitem{2016arXiv161100036D}
\bibinfo{author}{{DESI Collaboration}} \emph{et~al.}
\newblock \bibinfo{title}{{The DESI Experiment Part I: Science,Targeting, and Survey Design}} \bibinfo{pages}{arXiv:1611.00036} (\bibinfo{year}{2016}).

\bibitem{2024AJ....168...58D}
\bibinfo{author}{{DESI Collaboration}} \emph{et~al.}
\newblock \bibinfo{title}{{The Early Data Release of the Dark Energy Spectroscopic Instrument}}.
\newblock \emph{\bibinfo{journal}{Astron.\ J.}} \textbf{\bibinfo{volume}{168}}, \bibinfo{pages}{58} (\bibinfo{year}{2024}).

\bibitem{2012MNRAS.427..127B}
\bibinfo{author}{{Bressan}, A.} \emph{et~al.}
\newblock \bibinfo{title}{{PARSEC: stellar tracks and isochrones with the PAdova and TRieste Stellar Evolution Code}}.
\newblock \emph{\bibinfo{journal}{Mon.\ Not.\ R.\ Astron.\ Soc.}} \textbf{\bibinfo{volume}{427}}, \bibinfo{pages}{127--145} (\bibinfo{year}{2012}).

\end{thebibliography}

\clearpage
\section*{References} 
\begin{thebibliography}{10}
\makeatletter
\addtocounter{\@listctr}{19} 
\makeatother

\expandafter\ifx\csname url\endcsname\relax
  \def\url#1{\texttt{#1}}\fi
\expandafter\ifx\csname urlprefix\endcsname\relax\def\urlprefix{URL }\fi
\providecommand{\bibinfo}[2]{#2}
\providecommand{\eprint}[2][]{\url{#2}}

\bibitem{2001NuPhA.688..396T}
\bibinfo{author}{{Travaglio}, C.}, \bibinfo{author}{{Burkert}, A.} \& \bibinfo{author}{{Galli}, D.}
\newblock \bibinfo{title}{{Inhomogeneous chemical evolution of the Galactic halo}} \textbf{\bibinfo{volume}{688}}, \bibinfo{pages}{396--398} (\bibinfo{year}{2001}).

\bibitem{2017ApJ...849L...9T}
\bibinfo{author}{{Ting}, Y.-S.}, \bibinfo{author}{{Rix}, H.-W.}, \bibinfo{author}{{Conroy}, C.}, \bibinfo{author}{{Ho}, A. Y.~Q.} \& \bibinfo{author}{{Lin}, J.}
\newblock \bibinfo{title}{{Measuring 14 Elemental Abundances with R = 1800 LAMOST Spectra}}.
\newblock \emph{\bibinfo{journal}{Astrophys.\ J.\ Lett.}} \textbf{\bibinfo{volume}{849}}, \bibinfo{pages}{L9} (\bibinfo{year}{2017}).

\bibitem{2019ApJS..245...34X}
\bibinfo{author}{{Xiang}, M.} \emph{et~al.}
\newblock \bibinfo{title}{{Abundance Estimates for 16 Elements in 6 Million Stars from LAMOST DR5 Low-Resolution Spectra}}.
\newblock \emph{\bibinfo{journal}{Astrophys.\ J.\ Supplement}} \textbf{\bibinfo{volume}{245}}, \bibinfo{pages}{34} (\bibinfo{year}{2019}).

\bibitem{2015ApJS..216...29B}
\bibinfo{author}{{Bovy}, J.}
\newblock \bibinfo{title}{{galpy: A python Library for Galactic Dynamics}}.
\newblock \emph{\bibinfo{journal}{Astrophys.\ J.\ Supplement}} \textbf{\bibinfo{volume}{216}}, \bibinfo{pages}{29} (\bibinfo{year}{2015}).

\bibitem{2025arXiv251025876S}
\bibinfo{author}{{Sanders}, J.~L.}
\newblock \bibinfo{title}{{Chemical separation of stellar populations: analytic solutions for chemical evolution models with metallicity-dependent yields}}.
\newblock \emph{\bibinfo{journal}{arXiv e-prints}} \bibinfo{pages}{arXiv:2510.25876} (\bibinfo{year}{2025}).

\bibitem{2015ApJ...799..230H}
\bibinfo{author}{{Homma}, H.}, \bibinfo{author}{{Murayama}, T.}, \bibinfo{author}{{Kobayashi}, M. A.~R.} \& \bibinfo{author}{{Taniguchi}, Y.}
\newblock \bibinfo{title}{{A New Chemical Evolution Model for Dwarf Spheroidal Galaxies Based on Observed Long Star Formation Histories}}.
\newblock \emph{\bibinfo{journal}{Astrophys.\ J.}} \textbf{\bibinfo{volume}{799}}, \bibinfo{pages}{230} (\bibinfo{year}{2015}).

\bibitem{2015MNRAS.449..761U}
\bibinfo{author}{{Ural}, U.} \emph{et~al.}
\newblock \bibinfo{title}{{An inefficient dwarf: chemical abundances and the evolution of the Ursa Minor dwarf spheroidal galaxy}}.
\newblock \emph{\bibinfo{journal}{Mon.\ Not.\ R.\ Astron.\ Soc.}} \textbf{\bibinfo{volume}{449}}, \bibinfo{pages}{761--770} (\bibinfo{year}{2015}).

\bibitem{2009ARA&A..47..371T}
\bibinfo{author}{{Tolstoy}, E.}, \bibinfo{author}{{Hill}, V.} \& \bibinfo{author}{{Tosi}, M.}
\newblock \bibinfo{title}{{Star-Formation Histories, Abundances, and Kinematics of Dwarf Galaxies in the Local Group}}.
\newblock \emph{\bibinfo{journal}{Annual Review of Astron.\ Astrophys.}} \textbf{\bibinfo{volume}{47}}, \bibinfo{pages}{371--425} (\bibinfo{year}{2009}).

\bibitem{2019A&A...632A...4D}
\bibinfo{author}{{Di Matteo}, P.} \emph{et~al.}
\newblock \bibinfo{title}{{The Milky Way has no in-situ halo other than the heated thick disc. Composition of the stellar halo and age-dating the last significant merger with Gaia DR2 and APOGEE}}.
\newblock \emph{\bibinfo{journal}{Astron.\ Astrophys.}} \textbf{\bibinfo{volume}{632}}, \bibinfo{pages}{A4} (\bibinfo{year}{2019}).

\bibitem{kullback1951information}
\bibinfo{author}{Kullback, S.} \& \bibinfo{author}{Leibler, R.~A.}
\newblock \bibinfo{title}{On information and sufficiency}.
\newblock \emph{\bibinfo{journal}{Ann. Math. Statist.}} \textbf{\bibinfo{volume}{22}}, \bibinfo{pages}{79--86} (\bibinfo{year}{1951}).

\bibitem{2014Sci...344.1492R}
\bibinfo{author}{{Rodriguez}, A.} \& \bibinfo{author}{{Laio}, A.}
\newblock \bibinfo{title}{{Clustering by fast search and find of density peaks}} \textbf{\bibinfo{volume}{344}}, \bibinfo{pages}{1492--1496} (\bibinfo{year}{2014}).

\bibitem{2013ARA&A..51..457N}
\bibinfo{author}{{Nomoto}, K.}, \bibinfo{author}{{Kobayashi}, C.} \& \bibinfo{author}{{Tominaga}, N.}
\newblock \bibinfo{title}{{Nucleosynthesis in Stars and the Chemical Enrichment of Galaxies}}.
\newblock \emph{\bibinfo{journal}{Annual Review of Astron.\ Astrophys.}} \textbf{\bibinfo{volume}{51}}, \bibinfo{pages}{457--509} (\bibinfo{year}{2013}).

\bibitem{2016MNRAS.463.1518A}
\bibinfo{author}{{Amarsi}, A.~M.}, \bibinfo{author}{{Lind}, K.}, \bibinfo{author}{{Asplund}, M.}, \bibinfo{author}{{Barklem}, P.~S.} \& \bibinfo{author}{{Collet}, R.}
\newblock \bibinfo{title}{{Non-LTE line formation of Fe in late-type stars - III. 3D non-LTE analysis of metal-poor stars}}.
\newblock \emph{\bibinfo{journal}{Mon.\ Not.\ R.\ Astron.\ Soc.}} \textbf{\bibinfo{volume}{463}}, \bibinfo{pages}{1518--1533} (\bibinfo{year}{2016}).

\bibitem{2012A&A...538A..82R}
\bibinfo{author}{{Revaz}, Y.} \& \bibinfo{author}{{Jablonka}, P.}
\newblock \bibinfo{title}{{The dynamical and chemical evolution of dwarf spheroidal galaxies with GEAR}}.
\newblock \emph{\bibinfo{journal}{Astron.\ Astrophys.}} \textbf{\bibinfo{volume}{538}}, \bibinfo{pages}{A82} (\bibinfo{year}{2012}).

\bibitem{2014PASA...31...30K}
\bibinfo{author}{{Karakas}, A.~I.} \& \bibinfo{author}{{Lattanzio}, J.~C.}
\newblock \bibinfo{title}{{The Dawes Review 2: Nucleosynthesis and Stellar Yields of Low- and Intermediate-Mass Single Stars}} \textbf{\bibinfo{volume}{31}}, \bibinfo{pages}{e030} (\bibinfo{year}{2014}).

\bibitem{2012A&A...537A.146E}
\bibinfo{author}{{Ekstr{\"o}m}, S.} \emph{et~al.}
\newblock \bibinfo{title}{{Grids of stellar models with rotation. I. Models from 0.8 to 120 M$_{{\ensuremath{\odot}}}$ at solar metallicity (Z = 0.014)}}.
\newblock \emph{\bibinfo{journal}{Astron.\ Astrophys.}} \textbf{\bibinfo{volume}{537}}, \bibinfo{pages}{A146} (\bibinfo{year}{2012}).

\bibitem{2024IAUS..361..259T}
\bibinfo{author}{{Tsiatsiou}, S.}, \bibinfo{author}{{Georgy}, C.}, \bibinfo{author}{{Ekstr{\"o}m}, S.} \& \bibinfo{author}{{Meynet}, G.}
\newblock \bibinfo{title}{{Nitrogen production in population III stars}}.
\newblock In \bibinfo{editor}{{Mackey}, J.}, \bibinfo{editor}{{Vink}, J.~S.} \& \bibinfo{editor}{{St-Louis}, N.} (eds.) \emph{\bibinfo{booktitle}{Massive Stars Near and Far}}, vol. \bibinfo{volume}{361} of \emph{\bibinfo{series}{IAU Symposium}}, \bibinfo{pages}{259--260} (\bibinfo{year}{2024}).

\bibitem{2025arXiv250620436K}
\bibinfo{author}{{Kobayashi}, C.}
\newblock \bibinfo{title}{{Nucleosynthesis and the chemical enrichment of galaxies}}.
\newblock \emph{\bibinfo{journal}{arXiv e-prints}} \bibinfo{pages}{arXiv:2506.20436} (\bibinfo{year}{2025}).

\bibitem{2015MNRAS.453.1855M}
\bibinfo{author}{{Masseron}, T.} \& \bibinfo{author}{{Gilmore}, G.}
\newblock \bibinfo{title}{{Carbon, nitrogen and {\ensuremath{\alpha}}-element abundances determine the formation sequence of the Galactic thick and thin discs}}.
\newblock \emph{\bibinfo{journal}{Mon.\ Not.\ R.\ Astron.\ Soc.}} \textbf{\bibinfo{volume}{453}}, \bibinfo{pages}{1855--1866} (\bibinfo{year}{2015}).

\bibitem{2023ApJ...942...35C}
\bibinfo{author}{{Carrillo}, A.} \emph{et~al.}
\newblock \bibinfo{title}{{The Relationship between Age, Metallicity, and Abundances for Disk Stars in a Simulated Milky Way}}.
\newblock \emph{\bibinfo{journal}{Astrophys.\ J.}} \textbf{\bibinfo{volume}{942}}, \bibinfo{pages}{35} (\bibinfo{year}{2023}).

\bibitem{2011ApJ...729...16K}
\bibinfo{author}{{Kobayashi}, C.} \& \bibinfo{author}{{Nakasato}, N.}
\newblock \bibinfo{title}{{Chemodynamical Simulations of the Milky Way Galaxy}}.
\newblock \emph{\bibinfo{journal}{Astrophys.\ J.}} \textbf{\bibinfo{volume}{729}}, \bibinfo{pages}{16} (\bibinfo{year}{2011}).

\bibitem{2016A&A...589A.115S}
\bibinfo{author}{{Smiljanic}, R.} \emph{et~al.}
\newblock \bibinfo{title}{{The Gaia-ESO Survey: Sodium and aluminium abundances in giants and dwarfs. Implications for stellar and Galactic chemical evolution}}.
\newblock \emph{\bibinfo{journal}{Astron.\ Astrophys.}} \textbf{\bibinfo{volume}{589}}, \bibinfo{pages}{A115} (\bibinfo{year}{2016}).

\bibitem{2024A&A...682A.116N}
\bibinfo{author}{{Nissen}, P.~E.}, \bibinfo{author}{{Amarsi}, A.~M.}, \bibinfo{author}{{Sk{\'u}lad{\'o}ttir}, {\'A}.} \& \bibinfo{author}{{Schuster}, W.~J.}
\newblock \bibinfo{title}{{Abundances of iron-peak elements in accreted and in situ born Galactic halo stars}}.
\newblock \emph{\bibinfo{journal}{Astron.\ Astrophys.}} \textbf{\bibinfo{volume}{682}}, \bibinfo{pages}{A116} (\bibinfo{year}{2024}).

\bibitem{2025arXiv250506606E}
\bibinfo{author}{{Ernandes}, H.}, \bibinfo{author}{{Feuillet}, D.}, \bibinfo{author}{{Feltzing}, S.} \& \bibinfo{author}{{Sk{\'u}lad{\'o}ttir}, {\'A}.}
\newblock \bibinfo{title}{{Gaia-Sausage-Enceladus star formation history as revealed by detailed elemental abundances}}.
\newblock \emph{\bibinfo{journal}{arXiv e-prints}} \bibinfo{pages}{arXiv:2505.06606} (\bibinfo{year}{2025}).

\bibitem{2019MNRAS.488.1235M}
\bibinfo{author}{{Myeong}, G.~C.}, \bibinfo{author}{{Vasiliev}, E.}, \bibinfo{author}{{Iorio}, G.}, \bibinfo{author}{{Evans}, N.~W.} \& \bibinfo{author}{{Belokurov}, V.}
\newblock \bibinfo{title}{{Evidence for two early accretion events that built the Milky Way stellar halo}}.
\newblock \emph{\bibinfo{journal}{Mon.\ Not.\ R.\ Astron.\ Soc.}} \textbf{\bibinfo{volume}{488}}, \bibinfo{pages}{1235--1247} (\bibinfo{year}{2019}).

\bibitem{2019MNRAS.482.3426M}
\bibinfo{author}{{Mackereth}, J.~T.} \emph{et~al.}
\newblock \bibinfo{title}{{The origin of accreted stellar halo populations in the Milky Way using APOGEE, Gaia, and the EAGLE simulations}}.
\newblock \emph{\bibinfo{journal}{Mon.\ Not.\ R.\ Astron.\ Soc.}} \textbf{\bibinfo{volume}{482}}, \bibinfo{pages}{3426--3442} (\bibinfo{year}{2019}).

\bibitem{2020MNRAS.494.3880B}
\bibinfo{author}{{Belokurov}, V.} \emph{et~al.}
\newblock \bibinfo{title}{{The biggest splash}}.
\newblock \emph{\bibinfo{journal}{Mon.\ Not.\ R.\ Astron.\ Soc.}} \textbf{\bibinfo{volume}{494}}, \bibinfo{pages}{3880--3898} (\bibinfo{year}{2020}).

\end{thebibliography}
\end{document}